**RESEARCH ARTICLE**

# Title: A Tunable Terahertz Metamaterial Perfect Absorber Based on Vanadium Dioxide

**ZAHRA MADADI**[*]

Department of Electrical and Computer Engineering, Science and Research Branch, Islamic Azad University, Tehran, Iran.

**Abstract**

**Background:** In this paper, a new metamaterial perfect absorber based on vanadium dioxide ($VO_2$) is designed in the terahertz frequency range, which by changing the conductivity of $VO_2$, a tunability will be observed in the absorption spectrum of the device.

**Methods:** The proposed structure is simulated by the three-dimensional Finite Difference Time Domain (FDTD).

**Result:** Simulation results show that the absorption spectrum of the device in the normal incidence of a plane wave light in the range of 1 THz to 12 THz, has three polarization-insensitive resonance peaks whose amplitude changes for the different conductivities of $VO_2(\sigma)$ and shifts in a certain frequency range. For the VO2 ($\sigma$) in the metallic state ($\sigma = 2 \times 10^5$ S/m), as the refractive index of the analyte on top of the SiO2 spacer increases, a red shift is observed for all three peaks. The sensitivity ($S = \Delta\lambda/\Delta n$) of the first, second, and third peaks have been obtained as 69.73 GHz/RIU, 36.70 GHz/RIU, and 137.63 GHz/RIU, respectively. This device also having a tolerance of 6 degrees of incidence angle can provide a relatively stable absorption spectrum.

**Conclusion:** Features make this Absorber suitable for sensor applications. In future research, by placing the gold microparticle around the vanadium dioxide microparticle in a ring-shaped state or other designs, local plasmon resonances of gold can be used to improve the sensitivity of the proposed sensor.

## 1. INTRODUCTION

In the past decades, the metamaterial perfect absorbers (MPAs) have received extensive attention because of their widespread roles in sensing, energy harvesting, imaging, and stealth technology [1-3]. The most common basic structure of these absorbers is based on a metal-insulator-metal (MIM) configuration. So that there is a patterned metal array at the top and a flat continuous metal layer at the bottom of the structure, and these two layers are separated from each other by a dielectric spacer [4-6]. Noble metals such as gold (Au), silver (Ag), and aluminum (Al) are among the widely used materials in the structure of metamaterial absorbers, which can create strong resonances with almost perfect absorption through the stimulation of surface plasmon resonances [7-9]. Aluminum metal compared to gold and silver is more available and has a lower cost, but it has high reactivity with oxygen and produces weaker resonances [7]. Gold is a precious metal that is inert against most chemical reactions and has high durability [8], while silver metal is a little cheaper and can react with oxygen and sulfur, this leads to degradation of its plasmonic characteristics over time [9]. All of these metals have intrinsic absorption characteristics that limit the tunability of surface plasmon resonance. More precisely, it can be said that in these absorbers, depending on the material and their dimensions of layers, the periodicity of the upper metal array, the magnetic resonances will be stimulated along certain wavelengths, and the amount of absorption and the frequency position of the resonant peaks will be constant [10,11]. Therefore, the idea of designing tunable absorbers using semiconductors, liquid crystals, graphene, and phase change materials (PCM) has always been raised [3,12-14]. Vanadium dioxide ($VO_2$) is one of the most important phase change materials, whose crystal structure transforms from monoclinic (insulating phase) to tetragonal (metallic phase) at a critical temperature of 68 ºC [3,15]. Therefore, it is possible to achieve tunable dynamic characteristics such as the shift of the spectral position of the resonance peaks and their amplitude by exploiting the $VO_2$ material in metamaterial devices [15,16]. In this research, we propose a new vanadium dioxide-based metamaterial perfect absorber structure, which by changing its conductivity, a tunability will be observed in the absorption spectrum of the device. This paper is organized as follows; In section 2, the design of the absorber structure and its dimensions are reported. The simulation results obtained for the tunable metamaterial perfect absorber are described in Section 3. Finally, we conclude this research in section 4.

## 2. ABSORBER STRUCTURE

The unit cell of the proposed metamaterial perfect absorber structure is shown in Figure 1. As can be seen, the structure design consists of three layers: The bottom layer is made of gold (Au) and it has the same dimension in the x and y directions (Px=Py=30 µm) and a thickness of $t_1$=0.2 µm. The upper layer is a disc-shaped micro-particle with a cross in its middle circle, which is made of $VO_2$ with a thickness of $t_3$=0.08 µm so that the four quarter-circle holes have been created inside the disk. The radius of the outer circle of the disc is R=12 µm and the radius of the holes of the quarter-circle sectors is about r=3 µm. The width of the cross blades is d=4 µm. The middle layer of the structure is made of $SiO_2$ and it has a thickness of $t_2$=12 µm, which separates the upper and lower layers as a dielectric spacer.

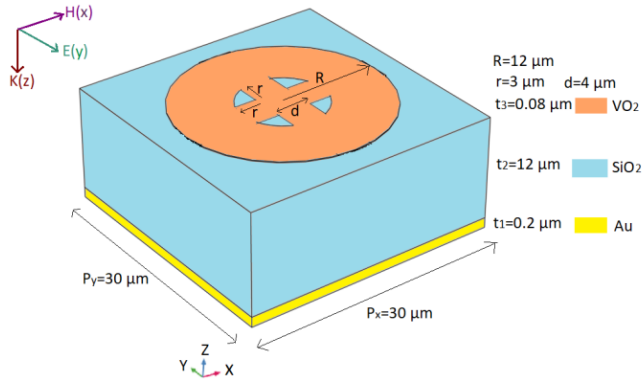

Fig. (1). Unit cell of the proposed tunable metamaterial perfect absorber.

## 3. SIMULATION RESULT

The proposed structure is simulated by the 3D Finite Difference Time Domain (FDTD) method. Light from a plane wave source with electric field polarization in the y-direction within the 1-12 THz frequency range incidences from top to bottom along the Z-axis. The simulation environment is limited from the upward and downward to the boundary condition of the perfect match layer (PML), and in the x and y-directions is limited by the periodic boundary condition. The periodicity of the unit cell in the x and y directions is $P_x=P_y$=30 µm. The dielectric constant of $VO_2$ in the terahertz band range can be expressed by the Drude model [3,17-19]:

$$\varepsilon(\omega) = \varepsilon_\infty - \frac{\omega_p^2(\sigma)}{\omega^2 + i\gamma\omega} \tag{1}$$

While the plasma frequency as a function of conductivity ($\omega_p(\sigma)$) is calculated as follows [3,17-19]:

$$\omega_p^2(\sigma) = \frac{\sigma}{\sigma_0} \omega_p^2(\sigma_0) \tag{2}$$

Where, $\varepsilon_\infty$ =12, $\gamma$ = 5.75×10$^{13}$ rad/s, $\sigma_0$ =3×10$^5$ S/m, and $\omega_p(\sigma_0)$= 1.4×10$^{15}$ rad/s. The conductivity of $VO_2$ ($\sigma$) in metallic state (tetragonal phase) and insulating state (monoclinic phase) are 2×10$^5$ S/m and 0 S/m, respectively [17-19]. The permittivity ($\varepsilon$) of silicon dioxide ($SiO_2$) is 3.8 and it has no losses in the desired frequency range. The conductivity of the bottom gold layer is considered 4.56×10$^7$ S/m.

To measure the absorption of the structure, firstly the values of the reflection coefficient ($S_{11}$) and transmission coefficient ($S_{21}$) are calculated. Then, the absorption coefficient will be calculated via the following equation [3,19-21]:

$$A(\omega) = 1 - R(\omega) - T(\omega)$$
$$= 1 - |S_{11}(\omega)|^2 - |S_{21}(\omega)|^2 \tag{3}$$

It should be noted that the Reflection spectrum ($R(\omega)$) is measured by a monitor at the top of the devices, and represents the intensity of power reflected from the overall structure. The Transmission spectrum ($T(\omega)$) is also measured by a monitor under structure [3,19-21]. In this design, the thickness of the bottom gold layer ($t_1$) is significantly greater than the penetration depth of the incident wavelengths, which acts as a full reflecting mirror and prevents light transmission from the structure ($T(\omega)$=0). Therefore, the absorption is obtained using the following relation [3,19-21]:

$$A(\omega) = 1 - R(\omega) = 1 - |S_{11}(\omega)|^2 \tag{4}$$

The absorption and reflection spectra associated with the proposed absorber for the different conductivities of $VO_2$ ($\sigma$) are computed and shown in Figure 2.

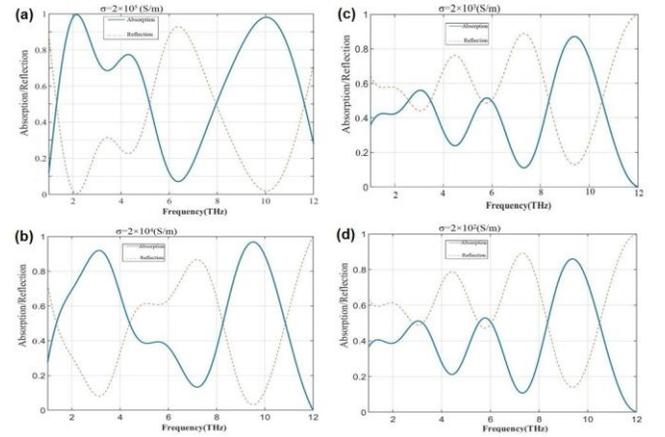

Fig. (2). The absorption and reflection spectra associated with the proposed tunable metamaterial perfect absorber for the different conductivities of $VO_2$ ($\sigma$).

In another step, to further understand the physical mechanism of absorption in the proposed device, an electric field analysis of the absorber for the VO2 ($\sigma$) in the metallic state (tetragonal phase) with electrical conductivity 2×10$^5$ S/m has been performed. For this purpose, the absolute value of the electric field distribution in z = 12.24 µm, x=0 µm, and -15 µm< y <15 µm has been measured at the absorption peaks frequencies of 2.13413 THz, 4.27027 THz, and 9.996 THz, and is shown in Figures 3(a), 3(B), and 3(c), respectively. According to the observations in three peaks, the absolute value of the electric field is maximum in the vicinity of the outer circle of the disk, in other words, the field is localized at the coordinates of y=+12.65 µm and y=-12.65 µm. Also, the distribution of the electric field is concentrated to some extent in the four quarter-circle holes inside the disk. By comparing three peaks and checking the exact numerical value of the field, it seems that for the peak is at a higher frequency, the absolute value of the electric field in the outer circle of the disc is decreased and instead the field localization is increased within the four quarter-circle holes.

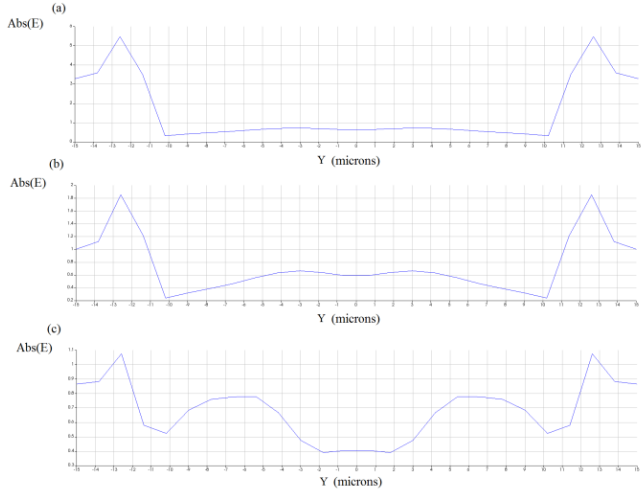

**Fig. (3).** The absolute value of electric field distribution for the $VO_2$ (σ) in the metallic state with electrical conductivity $2\times10^5$ S/m at the (a) first resonance frequency, (b) second resonance frequency, (c) third resonance frequency.

Once again, we consider the electrical conductivity of vanadium dioxide to be $2\times10^2$ S/m and measure the absolute value of the electric field distribution at the frequencies of the resonant peaks of the absorption spectrum, i.e. at 3.01502 THz, 5.75676 THz, and 9.34635 THz. The dimensions and location of the monitor displaying the field profile in the frequency domain are the same as before. Based on the results obtained and shown in Figure 4, for the first and second peaks, which are at lower frequencies, the absolute value of the electric field distribution in the vicinity of the outer circle of the disk has decreased compared to the previous state. In the third peak, there is also a slight decrease compared to the state of vanadium dioxide with electrical conductivity of $2\times10^5$ S/m. In other words, as the electrical conductivity of vanadium dioxide changes step by step from the metal phase to the insulating phase, the absolute value of the electric field distribution in the vicinity of the disc-shaped micro-particle and cross in the middle circle decreases and thus the amplitude of the resonance peaks in the absorption spectrum decreases.

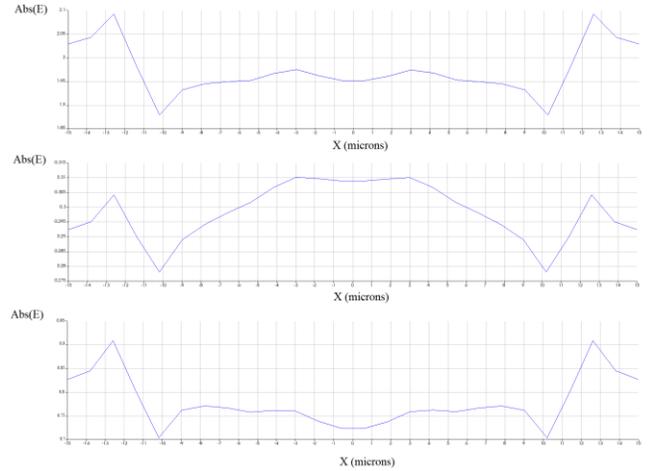

**Fig. (4).** The absolute value of electric field distribution for the $VO_2$ (σ) in metallic state with electrical conductivity $2\times10^2$ S/m at the (a) first resonance frequency, (b) second resonance frequency, (c) third resonance frequency.

It should be noted that due to the symmetry of the proposed absorber structure, the absorption spectrum of the device under normal incidence of the plane wave light is insensitive to polarization. In order to investigate the sensitivity of the absorption spectrum to the incidence angle in the case where vanadium dioxide electrical conductivity is $2\times10^5$ S/m, we increased the radiation angle from 0 (normal incidence) with steps of 2 degrees. After obtaining the results, which can be seen in Figure 5, the absorption spectrum was found to be almost constant by increasing the radiation angle up to 6 degrees. In better words, the proposed structure having a tolerance of 6 degrees in normal incidence can provide a relatively stable absorption spectrum.

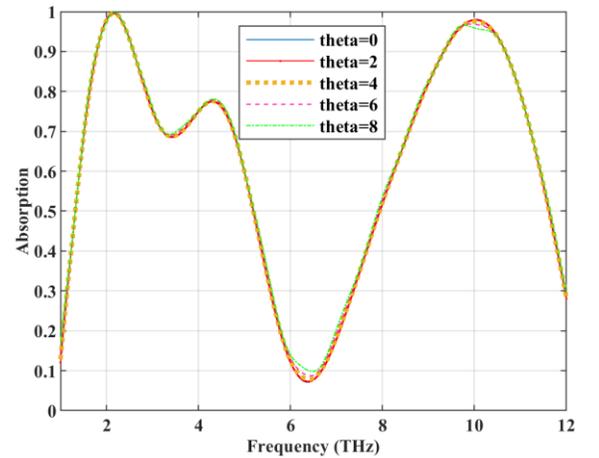

**Fig. (5).** The absorption spectrum of the proposed perfect absorber for the different incidence angles.

In another step, in order to test the sensing performance of the proposed absorber for the $VO_2$ (σ) in the metallic state (σ =$2\times10^5$ S/m), we place a 200 nm thick analyte layer on top of the $SiO_2$ spacer. According to the observations (Figure 6), as the refractive index of the analyte increases from n=1 to

n=6, a red shift is observed for all three peaks in the investigated spectral range. Thus, the sensitivity (S=Δλ/Δn) for the first, second, and third peaks have been obtained as 69.73 GHz/RIU, 36.70 GHz/RIU, and 137.63 GHz/RIU, respectively. In future research, by placing the gold microparticle around the vanadium dioxide microparticle in a ring-shaped state or other designs, local plasmon resonances of gold can be used to improve the sensitivity of the proposed absorber sensor.

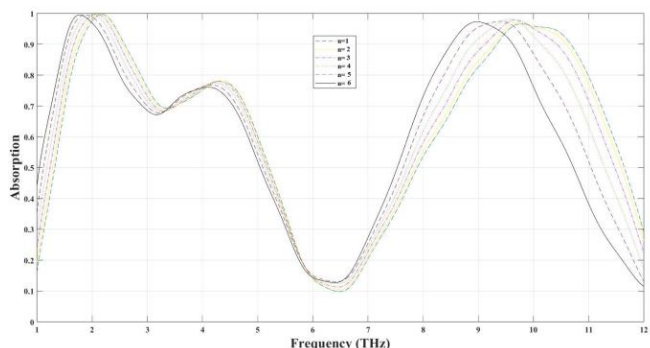

**Fig. (6).** The absorption spectrum of the proposed perfect absorber for the different analyte refractive indices (n=1,2,3,4,5,6).

Finally, it is necessary to note that a lot of research has been done in the field of the design of plasmonic absorbers and their use as sensors. As an example, the following can be mentioned. In the structure proposed by Hongsen Zhang et al., a quad-band plasmonic perfect absorber has been designed and the sensitivity of its peaks has been obtained as 4367 nm/RIU, 2162 nm/RIU, 1059 nm/RIU, and 908 nm/RIU respectively [22]. Zhiren Li et al., have also proposed a dual-band terahertz perfect absorber based on the all-dielectric metamaterial composed of the vertical-square-split-ring structure InSb array whose peaks sensitivity have been obtained as 1.3 THz/RIU and 1 THz/RIU  under T= 295 K [23]. J. Zhao et al. have proposed a temperature-tunable terahertz perfect absorber based on an all-dielectric Strontium Titanate resonator structure used as a temperature sensor with a sensitivity of 0.37 GHz/K [24]. In the research of Yongzhi Cheng et al., a six-band terahertz perfect metasurfce absorber based on a single circular-split-ring resonator structure is proposed for biological sensing, at material detection at THz regions [25]. In another structure proposed by Yongzhi Cheng et al., a narrowband perfect metasurface absorber based on a micro-ring-shaped GaAs array as a sensor has achieved a sensitivity of 1.45 THz/RIU [26]. The important point is that in none of the research mentioned above, the tunable materials were not used in the absorber structure. So after the construction of the proposed devices according to the calculated dimensions of structure components, it is not possible for resonance peaks to shift at the desired frequency. Therefore, the important advantage of our proposed structure is the tunability of the output absorption spectrum of the device by changing the conductivity of vanadium dioxide.

## 4. CONCLUSION

In this paper, a tunable metamaterial perfect absorber based on vanadium dioxide has been presented in the terahertz frequency range. The proposed absorber is configured as metal-insulator-metal so that in the unit cell of the structure, a gold mirror layer is placed at the bottom, which blocks the transmission of light from under the structure. An insulating spacer of $SiO_2$ is layered on this gold substrate, and on top of it, a disk-shaped micro-particle of $VO_2$ with a cross in its middle circle is placed, whose crystal structure can be adjusted between the metallic and insulating phases. When it is in the metal phase, the proposed structure in the MIM configuration achieves high absorption, and by changing its crystal structure, the tunability of the absorption spectrum is achieved. Simulation results showed that the absorption spectrum of the proposed device is insensitive to polarization in normal incidence and even with having a tolerance of 6 degrees of incidence angle can provide a relatively stable absorption spectrum.

**CONSENT FOR PUBLICATION**

Not applicable.

**FUNDING**

None.

**AVAILABILITY OF DATA AND MATERIALS**

Not applicable.

**CONFLICT OF INTEREST**

None.

**ACKNOWLEDGEMENTS**

None declared.